\documentstyle[epsfig]{aipproc}

\begin{document}
\title{Prospects of observing pulsed radiation from
gamma-ray pulsars with H.E.S.S.}

\author{\bf O. C. de Jager$^1$, A. Konopelko$^2$, B.C. Raubenheimer$^1$
 and B. Visser$^1$}
\address{$^1$Unit for Space Physics, Potchefstroom University, Potchefstroom,
2520, South Africa\\ 
$^{2}$ Max Planck Institut f\"ur Kernphysik, Postfach 103980, D-69029
Heidelberg, Germany}

\maketitle

\begin{abstract}
Observations and theoretical studies have demonstrated that the pulsed
spectra of all gamma-ray pulsars terminate at energies below 
a few hundred GeV. In most cases we expect this cutoff energy $E_o$ to be
around 10 GeV.
Only with next-generation ground-based $\gamma$-ray telescopes, which
are expected to have non-zero trigger probabilities near 10 GeV, can
we expect to detect pulsations. The large $\gamma$-ray fluxes
below $E_o$, together with the associated sharp pulse profiles,
compensate for the lack of imaging capability
near threshold. For H.E.S.S. we find that the pulsed component of
PSR B1706-44 should be detectable near threshold, whereas the unidentified
GeV EGRET sources should be detectable if the superexponential cutoff energy
is larger than $\sim 30$ GeV for 
relatively hard pulsar photon spectra ($\sim E^{-1.5}$).  
\end{abstract}
\subsection{Introduction}
Whereas gamma-ray pulsars are known to count amongst the brightest
sources in the 1 - 30 GeV range (Lamb \& Macomb, 1997), only their plerions,
supernova shells and extragalactic sources appear to be
visible at TeV energies. This is not unexpected, since pulsed $\gamma$-rays
are known to be created in strong magnetic fields and magnetic pair
production results in a superexponential cutoff above a characteristic
energy $E_o$. Simulations of pair cascades
have shown that $E_o$ is usually in the 10 GeV region
(Daugherty \& Harding, 1996), but this cutoff depends on several
parameters, such as the altitude of $\gamma$-ray production
above the polar cap, the 
observer's viewing angle relative to the spin axis, and the 
magnetic inclination angle relative to the spin axis. A
$\gamma$-ray/e$^{+/-}$ cascade develops and those photons which
escape pair creation (with $E<E_o$) are observable, resulting in a
pulsed spectrum which is typically harder than $E^{-2}$ 
for $E<E_o$. Based on this consideration, Nel \& de Jager
(1995) modelled the high energy $\gamma$-ray pulsed spectra of pulsars
as

\begin{equation}
dN_{\gamma}/dE=k(E/E_n)^{-g}exp(-(E/E_o)^b). \noindent
\end{equation}

Whereas pulsar photon spectral indices between $g=1.4$ and 2.1
are observed, harder spectra are theoretically possible
(A.K. Harding, 2000, personal communication to O.C. de Jager).
The constant $k$ represents the monochromatic flux at the normalising
energy $E_n\ll E_o$. We will normalise spectra at $E_n=1$ GeV. 

In the case of the outergap model for pulsars (Cheng, Ho, \& Ruderman
1996), $\gamma$-ray production is expected to occur near the
pulsar light cylinder, and the cutoff is expected to result from
energetics arguments, rather than from magnetic pair production. In this
case a larger $E_o$ may be observable. Ground-based TeV $\gamma$-ray 
observations however provide firm upper limits on $E_o$. 
(Nel et al. (1993) gave a detailed discussion on this topic; see  
also Catanese \& Weekes 1999.)


\subsection{Gamma-Ray Pulsar Spectral Parameters above 1 GeV}
Table 1 shows
the parameter results of a fit to the total pulsed spectra of the six
brightest EGRET
$\gamma$-ray pulsars. These spectral parameters reproduce the
EGRET flux up to 30 GeV, and are consistent with the TeV pulsed limits. They 
also reproduce the GeV source catalog flux (Lamb \& Macomb 1997).
In the case of Vela and Geminga the cutoffs are well defined by the
EGRET data and the errors on $E_o$ are relatively small ($\sim 20\%$).
In the case of Crab and PSR
B1055-52, some evidence of a turnover is seen in the spectra above 10 GeV,
although it is difficult to obtain reliable measures of $E_o$ and $b$. In the
case of PSR B1951+32 and PSR B1706-44 we see no evidence of a turnover
up to 30 GeV, and a minimum value of $E_o=40$ GeV (consistent with EGRET)
was selected. This value is conservative with respect to the H.E.S.S.
response. For those cases where $E_o$ is not well defined, we have selected
$b=2$ (a value typical for a spectrum attenuated by magnetic pair
production) to give conservative H.E.S.S. rates.

Using the H.E.S.S. collection area vs. energy $A(E)$ for any 2-telescope
triggers (Konopelko 2000), 
we were able to calculate the expected rates $R_p$ for pulsed $\gamma$-rays
by integrating the product of $A(E)dN_{\gamma}/dE$ over all energies.
The results for the six EGRET pulsars are shown in Table 1 (indicated
by ``$R_p$''). It is clear that the rate for PSR B1706-44 is the
largest of all pulsars if $E_o$ is not smaller than 40 GeV.
     
\begin{table}
\begin{center}
\caption{Gamma-ray spectral parameters above 1 GeV and corresponding
H.E.S.S. rates and observation time for detection. Spectral references
from Macomb \& Gehrels (1999).}
\begin{tabular}{lccccccc}
Object & $k$ ($\times 10^{-8}$) & $g$ & $E_o$ & $b$ & $F(>1\;{\rm GeV})$ 
& $R_p$ & $T$ (10-hour \\
&(cm$^{-2}$s$^{-1}$GeV$^{-1}$) & & (GeV) & & (cm$^{-2}$s$^{-1}$) 
& (hour$^{-1}$) & days)\\
\hline
Crab         & 24.0   & 2.08 & 30  & 2   & 22   & 100  & 3 \\
Vela         & 138    & 1.62 & 8.0 & 1.7 & 148  & 8    & 400 \\
Geminga      & 73.0   & 1.42 & 5.0 & 2.2 & 76   & $\ll 1$& - \\
PSR B1951+32 & 3.80   & 1.74 & 40  & 2   & 4.9  & 180  & 1 \\
PSR B1055-52 & 4.00   & 1.80 & 20  & 2   & 4.5  & 8    & 420 \\
PSR B1706-44 & 20.5   & 2.10 & 40  & 2   & 20   & 240  & 1 \\
\hline
\end{tabular} 
\end{center} 
\end{table}

\subsection{H.E.S.S. Sensitivity for Pulsed $\gamma$-Ray Mission}
It was shown by de Jager, Swanepoel \& Raubenheimer (1987) and
de Jager (1994) that the basic scaling parameter for any test for
uniformity on the circle (given a test period) is given by
$x=p\sqrt{n}$, where $p=R_p/(R_b+R_p)$ is the pulsed fraction, with
$R_p$ the pulsed rate and $R_b$ the background rate. The total number of
events is given by $N=(R_p+R_b)T$, with
$T$ the observation time. In this case the test statistic
for uniformity for the general Beran (1969) class of tests is given by
$B=x^2\Phi_B +c$, where $\Phi_B$ is derived from the intrinsic pulse profile,
and c is the noise term. It was shown by Thompson (2000)
that the pulse profiles above 5 GeV consist mostly of a 
single narrow peak, and it can be shown that $\Phi_B=5.8$ for a
5\% FWHM (single peak), if $B$ is taken as the
$Z^2_m$ test statistic with $m=10$ harmonics (see e.g. de Jager,
Swanepoel \& Raubenheimer 1987). In this case $c=20$.

A value of $x=3$ would introduce a $\sim 3\sigma$ DC excess in a 
spatial analysis, but assuming that we have no imaging capability
for $E_o$ near the detection threshold, we have to rely on a timing
analysis, which would give $Z^2_{10}\sim 73$, or a chance
probability of $7\times 10^{-8}$ if the period is known, but 0.03
after multiplying with the number of trials for a 6 hour observation
if searching for periods as short as 50 ms. A confirming run
(e.g. on a second night) 
should always be made to see if one of the few most 
significant periods from the previous run have repeated 
itself - in this case at the $\sim 10^{-7}$ level.

Using an additional topological software trigger, and
selecting events by image size and angular shape, we were
able to reject $\sim 99.2\%$ of the 
triggered background events, while retaining 95\% of the source events. 
From a total background rate of 1 kHz (Konopelko 2000),
we get $R_b=8$ Hz. This allows us to calculate detection
sensitivities for periodicities:
   
From the GeV source catalogue, we find that
the galactic unidentified EGRET source 
(some may be pulsars - Lamb \& Macomb 1997) fluxes range from 
$F(>1\;{\rm GeV})$ = 1 to 25 $\times 10^{-8}$ cm$^{-2}$s$^{-1}$.
Figures 1 and 2 give the H.E.S.S. sensitivity for a wide
range of possible pulsar photon spectral indices 
between 1 and 2, and requiring a marginal detection within $T=3$ to
6 hours (assuming a minimum ``DC significance'' of $x=3$):
Figures 1 and 2 respectively show $E_o$ and $T$ vs $k$, with the
latter within the EGRET range as discussed above.
Table 1 also shows $T$ calculated in the same way, but 
assuming the spectral parameters of individual pulsars.
\subsection{Conclusions}
It is clear that H.E.S.S. can only detect pulsars if $E_o$
exceeds $\sim 30$ GeV. Even weak EGRET sources may be detectable
if the spectra are as hard as $E^{-1}$, provided that $E_o$
exceeds the levels prescribed by Figure 1.
PSR B1706-44 (for which $E_o$ is known to be at least as large
as $\sim 40$ GeV) should be a H.E.S.S. candidate and
other similar pulsars (such as PSR B1951+32) may be similarly
detectable within one night. If one cannot detect a clear signal
within a single night, an exact timing solution would be
required to do a coherent analysis over a long period of time.

Whereas we have addressed the conservative polar cap model, any
outergap component is expected to give a large value for $E_o$
(which is no challenge for H.E.S.S.), but $k$ may be small for
such pulsars. This will be treated in a separate paper.
 
\begin{figure}
\vspace{-1.8cm}
\centerline{
\epsfxsize=8.5cm
\hspace*{1.0cm}\epsffile{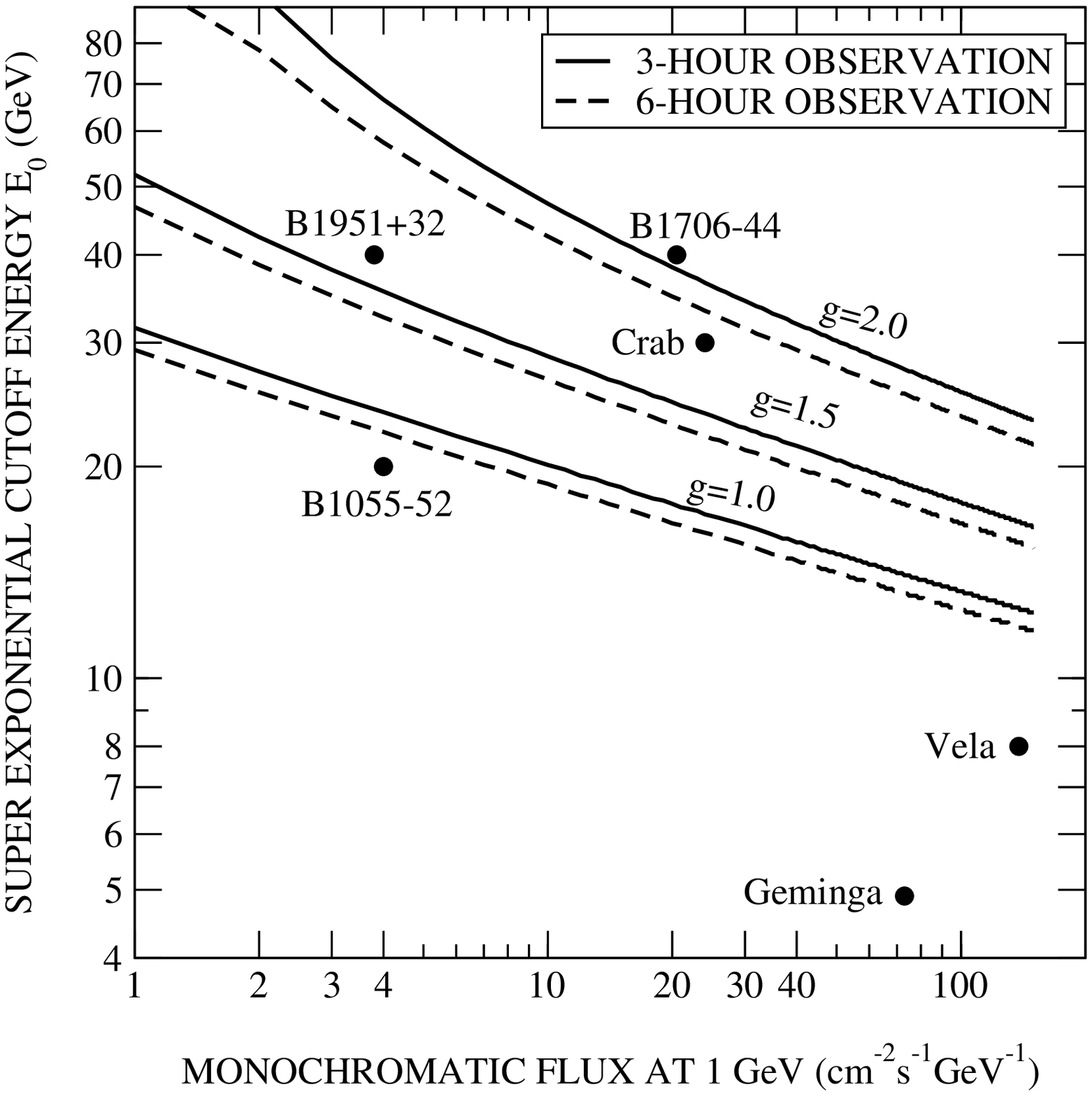}
\epsfxsize=8.5cm
\hspace*{-0.4cm}\epsfbox{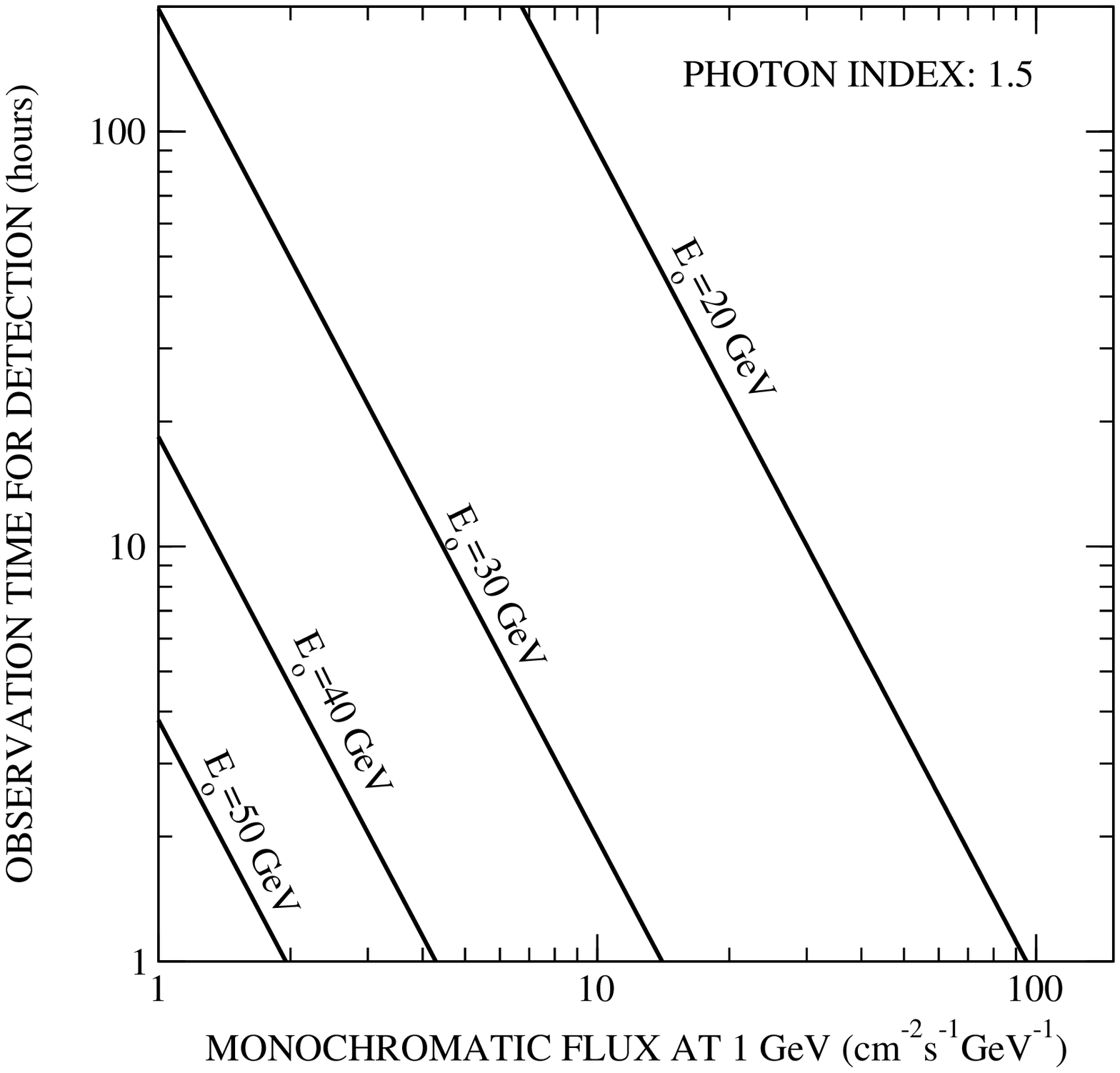} 
}
\vspace{-3cm}
\caption{{\bf Figure 1} (Left panel): Parameter space 
($E_o$ vs $k$) for the detection of unknown pulsars within one night 
with H.E.S.S. using a timing analysis approach, and assuming $x=3$.
The three
curves represent (from bottom to top) photon spectral indices of 1, 1.5
and 2.0. The solid line is for 3 hours of continuous observation, whereas
the dashed lines (for the same set of spectral indices) represent a six-hour
run.
{\bf Figure 2} (right panel): The observation time required to detect a pulsar
as a function of $k$ for a photon index of 1.5 and $E_o$ as shown
(also for $x=3$).}

\end{figure}


\subsection*{References}
\def\rf{\noindent\hangafter=1\hangindent=1truecm}

\rf 
Beran, R.J. 1969, Ann. Math. Statist., 40, 1196.

\rf
Catanese, M. \& Weekes, T.C. 1999, PASP,111(764), 1193.

\rf
Cheng, K.S., Ho, C. \& Ruderman, M.A. 1986, ApJ, 300, 500.

\rf
Daugherty, J.K. \& Harding, A.K. 1996, ApJ, 458, 278.

\rf
de Jager, O.C., Swanepoel, J.W.H. \& Raubenheimer, B.C. 1986, A\&A, 170, 187.

\rf
de Jager, O.C. 1994, ApJ, 436, 239.


\rf
Konopelko, A. 2000, these proceedings.

\rf
Lamb, R.C. \& Macomb, D.J., 1997, ApJ, 488, 872
 
\rf
Macomb, D.J. \& Gehrels, N. 1999, ApJ Suppl, 120, 335. 

\rf 
Nel, H.I. et al. 1993, ApJ, 418, 836.

\rf
Nel, H.I. \& de Jager, O.C. 1995, Astr. Space Science, 230, 299.

\rf
Thompson, D.J. 2000, these proceedings.

\end{document}